\documentclass[12pt,nofootinbib]{revtex4}
\usepackage{amsmath,amsfonts,amssymb} 
\usepackage[pdftex]{graphicx} 
\usepackage{setspace}
\usepackage[margin=0.8in, paperwidth=8in, paperheight=11in]{geometry}
\usepackage{bbm}
\begin{document}
\title{Spontaneous emergence of free-space optical and atomic patterns}
\author{Bonnie L. Schmittberger$^1$ and Daniel J. Gauthier$^{1,2}$}
\affiliation{$^1$Duke University, Department of Physics and Fitzpatrick Institute for Photonics, Box 90305, Durham, North Carolina 27708, USA}
\affiliation{$^2$The Ohio State University, Department of Physics, 191 West Woodruff Ave., Columbus, OH 43210 USA}
\begin{abstract}
The spontaneous formation of patterns in dynamical systems is a rich phenomenon that transcends scientific boundaries. Here, we report our observation of coupled optical-atomic pattern formation, which results in the creation of self-organized, multimode structures in free-space laser-driven cold atoms. We show that this process gives rise to spontaneous three-dimensional Sisyphus cooling even at very low light intensities and the emergence of self-organized atomic structures on both sub- and super-wavelength scales.
\end{abstract}
\maketitle

\vspace{10mm}

\section{Introduction}
Optical lattices are a useful experimental tool for studying the interaction between light and ultracold atoms. The lattice creates a dipole force that results in spatial organization of the atoms. Spatially organized atoms can be used to simulate quantum systems~\cite{Blochreview} and to study novel states of matter~\cite{MottInsulator}. These applications often require only a small number of atoms per lattice site, but new physics is expected when there are many atoms per site~\cite{GopLevGoldNatPhys}.

In the high-atom-number regime, the light scattered by the atoms can be substantial and affect the atoms' center-of-mass degrees-of-freedom. Early experiments exploring this regime used single-mode optical cavities to enhance the light-atom coupling strength~\cite{Ritsch}. Alternatively, we have shown theoretically that strong light-atom coupling can be achieved in free space by allowing many sub-Doppler-cooled atoms to spatially organize in the intensity maxima of an optical lattice~\cite{PhysRevA.90.013813}. Free-space systems are advantageous for many reasons, including experimental simplicity and access to an intrinsically multimode system~\cite{optopatterns,PhysRevA.86.013823}. Strong light-atom interactions in multimode geometries allow access to different physics, such as continuous symmetry-breaking and the realization of spin glasses, for example~\cite{GopLevGoldNatPhys}.

In this article, we realize enhanced light-atom interactions in a multimode, free-space cloud of ultracold atoms driven by counterpropagating optical fields. Above a threshold value of the nonlinear refractive index, denoted $n_{\text{NL}}$, we observe an instability that simultaneously generates new optical fields and new real-space gratings of atoms, which we refer to as optical/atomic pattern formation. The optical and atomic patterns enhance each other because the generated fields give rise to atomic cooling and real-space self-organization, which in turn give rise to increased optical scattering. This results in a runaway process, known as an absolute instability, where an initial atom/field fluctuation triggers macroscopic pattern formation.

While other cold-atom systems have observed the spontaneous emergence of multimode optical fields~\cite{optopatterns,PhysRevA.86.013823}, they have not directly measured a corresponding self-organization of the atoms. Here, we demonstrate real-space self-organization of atoms in a multimode system using parametric resonance techniques, which allows us to both verify self-organization and quantitatively measure the characteristics of the self-organized atoms. We observe that the self-organized patterns exhibit continuous symmetry-breaking as well as long-range ordering on both sub- and super-wavelength scales. We also find that the atoms spontaneously cool in 3D despite only applying fields in one dimension. Three-dimensional Sisyphus cooling has previously been studied using 3D applied optical lattices to realize longer coherence times~\cite{PhysRevLett.78.630}, but it has not yet been observed to occur spontaneously, as it does in our system. Finally, we observe pattern formation at ultra-low intensities, which allows for studies of low-light-level nonlinear optics and multimode non-equilibrium phenomena and provides a novel material for studying the spontaneous formation of patterns and the nature of absolute instabilities.

\section{Methods}
Our free-space experiment consists of an elongated sample of sub-Doppler-cooled $^{87}$Rb atoms created in a magneto-optical trap (MOT) of length $L=3\text{ cm}$ and diameter $w{\sim}400\text{ }\mu\text{m}$ that are initially cooled to a temperature $T\simeq30\text{ }\mu$K~\cite{Greenberg:07}, well below the Doppler temperature $T_D=146\text{ }\mu\text{K}$. The typical on-resonance optical depths are $50-100$ along the long trap axis with an average density of $n_a\sim10^{10}\text{ atoms/cm}^3$. After loading the MOT for 97 ms, we apply counterpropagating optical fields (wavelength $\lambda\sim780\text{ nm}$, $1/e^2$ beam waist $\sim410\text{ }\mu\text{m}$) in a lin$\perp$lin polarization configuration along the long axis of the atomic cloud for 3 ms, which form an imposed 1D optical lattice with ${\sim}200$ atoms per site. The total applied electric field is given by $\vec{E}(z,t)=\vec{E}_0(z)e^{-i\omega t}+\text{c.c.}$, where $\vec{E}_0(z)=F(z)e^{ikz}\hat{x}+iB(z)e^{-ikz}\hat{y}$, $k$ is the vacuum wavenumber, $\omega$ is the frequency, and $i$ is an arbitrary phase chosen to set the polarization to $\hat{\sigma}^-$ at $z=0$. This electric field gives rise to a spatially-modulated AC Stark shift in the atomic energy states~\cite{Metcalf}. When the pump fields have a detuning $\Delta=\omega-\omega_{eg}$ below the atomic resonance frequency $\omega_{eg}$ ($\Delta<0$), the atoms undergo initial Sisyphus cooling along the long $\hat{z}$-axis~\cite{SisyphusCastinmain}. After $\sim30\text{ }\mu\text{s}$, the mean temperature of the gas along $\hat{z}$ is cooled to $T_z\simeq2-3$ $\mu$K, while the mean temperature in the radial directions is still $T_{\text{rad}}\simeq30\text{ }\mu$K~\cite{Greenberg:11}. We typically use $\Delta=-4\Gamma$ to $-10\Gamma$, where $\Gamma=2\pi\times(6\text{ MHz})$ is the natural linewidth of the $5^2\text{S}_{1/2}\text{ }(\text{F}=2)\rightarrow5^2\text{P}_{3/2}\text{ }(\text{F}=3)$ atomic transition, which is close enough to the resonance to observe pattern formation, but far enough to avoid substantial absorption~\cite{SchmittbergerPatternTheory}. This regime is atypical for lattice experiments, where typically $|\Delta|\gg100~\Gamma$~\cite{PhysRevLett.106.223903}, and atom-field interaction strengths are comparatively weak.
\begin{figure}
\begin{center}
 \includegraphics[scale=0.4]{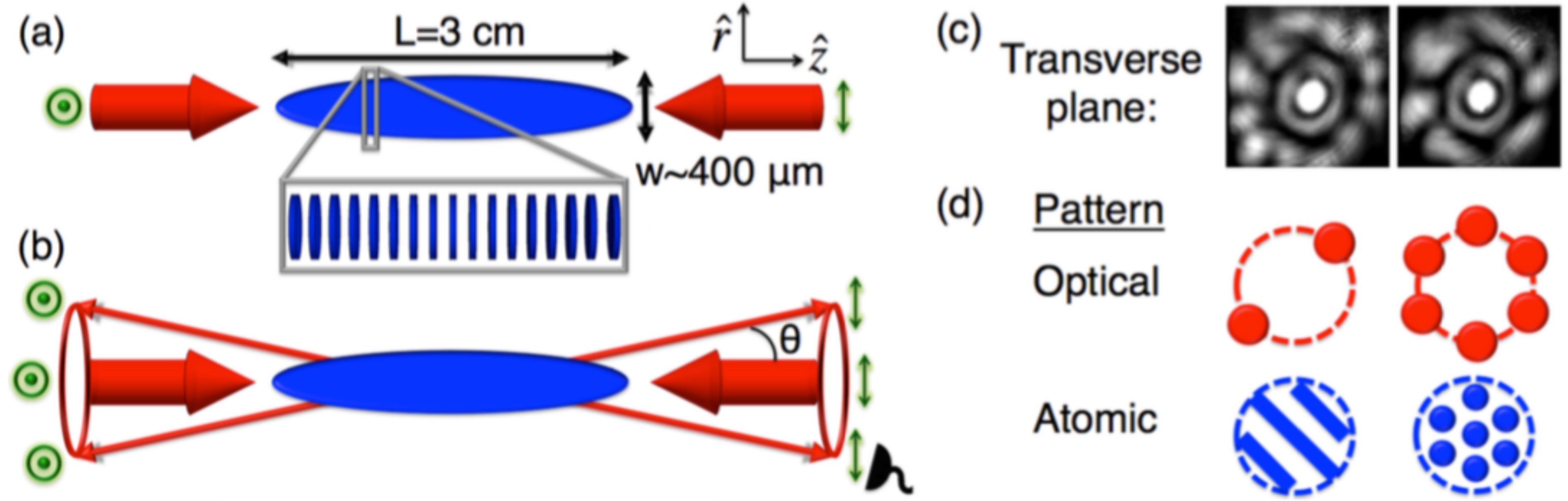}
 \caption{(a) Counterpropagating optical fields with a lin$\perp$lin polarization configuration (shown with green circles/arrows) drive an elliptical cloud of cold atoms along the $\hat{z}$-axis. The atoms spatially organize into pancake-shaped structures, depicted in the rectangle. (b) Instability-generated optical fields propagate along a cone surrounding the pump fields. The placement of the detector is shown as the black semi-circle. (c) Far-field examples of a 12-spot and a 6-spot transverse optical pattern imaged along $\hat{z}$ in contiguous experimental runs (100 ms apart), where the MOT was reloaded between each image, with $I/I_{\text{sat}}=0.4$, and the optical depth is $62$. The central spot is bleed-through pump light. The small, central ring closely surrounding the pump spot is a pump beam reshaping effect. (d) Simulations of the atomic pattern formed in a pancake due to two- and six-spot optical patterns.}
 \label{multimodepapersetupfig4}
  \end{center}
 \end{figure}

The spatially-varying AC Stark shift of the applied optical lattice forms two superimposed dipole potentials, the total of which goes as $U(z)=U^+(z)+U^-(z)$~\cite{SisyphusCastinmain}, where
\begin{equation}
U^{\pm}(z)=U_0\left[1\mp\frac{1}{2}\text{cos}(2k^\prime z)\right]
\label{dipolepotentialterms}
\end{equation}
correspond to the dipole potentials for the $m_J=\pm1/2$ ground states in a $J=1/2\rightarrow J^\prime=3/2$ atomic transition. This simplified fine-structure model provides a good qualitative understanding of our experiment because most (${\sim}90\%$) of the atoms are tightly bunched at regions of pure $\hat{\sigma}^+$/$\hat{\sigma}^-$ polarization and pumped into the $F=2$, $m_F=\pm2$ ground states~\cite{Greenberg:11}. The wavenumber of the pump fields propagating through the atomic cloud is $k^\prime=kn$ for atoms with an effective index of refraction $n$, defined below. The maximum well depth in the low-intensity limit is given by
\begin{equation}
U_0=\frac{\hbar\Delta I}{I_\text{sat}},
\label{Udippump}
\end{equation}
where $I=2\epsilon_0c\left|E_0(z)\right|^2$ is the total intensity, $I_{\text{sat}}=\hbar^2\Gamma^2\epsilon_0c/[2|\mu|^2C^2(1+4\Delta^2/\Gamma^2)]$ is the off-resonant saturation intensity~\cite{Boyd}, $\epsilon_0$ is the permittivity of free space, $c$ is the speed of light, $|\mu|$ is the dipole moment magnitude for $m_J=\pm1/2\rightarrow m_{J^\prime}=\pm3/2$ transitions, and $C^2=2/3$ is the difference between the square of the Clebsch-Gordon coefficients for the $m_J=\pm1/2\rightarrow m_{J^\prime}=\pm3/2$ and $m_J=\pm1/2\rightarrow m_{J^\prime}=\mp1/2$ transitions~\cite{SchmittbergerPatternTheory}.

The dipole potential gives rise to an imposed density distribution
\begin{equation}
\eta(z)=\frac{n_a}{2}\left[\sum_{j=-\infty}^{\infty}\eta_j^+e^{2i(k^\prime z-\pi/2)*j}+\sum_{j=-\infty}^{\infty}\eta_j^-e^{2ik^\prime z*j}\right],
\label{densitydist}
\end{equation}
where the grating wavevector for bunched atoms in a given spin state is $\vec{g}_p=\pm2k^\prime\hat{z}$. The two sums in Eq.~\ref{densitydist} represent atomic bunching at locations of pure $\hat{\sigma}^\pm$ field polarizations, where neighboring pancakes have opposite spins. For a gas in thermal equilibrium~\cite{SchmittbergerPatternTheory}, the Fourier coefficients appearing in Eq.~\ref{densitydist} are given by
\begin{equation}
\eta_j^\pm=\frac{I_j[-\zeta]}{I_0[-\zeta]},
\label{FCSisyphus}
\end{equation}
where $I_j$ refers to a modified Bessel function of the first kind of order $j$, and $\zeta=[(\Delta/\Gamma)(I/I_{\text{sat}})]/(T_z/T_D)$ is the ratio of $U_0$ to the atomic thermal energy. 

The first-order Fourier coefficients $\eta_1^{\pm}$ provide a measure of atomic bunching, where $\eta_1^{\pm}=0$ ($\zeta=0$) indicate a homogeneous gas, and $|\eta_1^{\pm}|=1$ ($|\zeta|\rightarrow\infty$) indicate maximum bunching, \textit{i.e.}, infinitely thin sheets of atoms~\cite{PhysRevA.52.1394}. By increasing $\eta_1^{\pm}$ for $\Delta<0$, the light-atom coupling strength is enhanced because atoms bunch tightly at the intensity maxima of the lattice. As can be seen from Eq.~\ref{FCSisyphus}, atomic bunching provides a new mechanism to achieve enhanced nonlinear light-atom interactions even for $I/I_{\text{sat}}\ll1$ by using small $|\Delta|$, which increases the dipole potential well depth, and by using small $T_z$ (\textit{e.g.}, via Sisyphus cooling)~\cite{PhysRevA.90.013813}. This bunching-induced nonlinearity is the primary mechanism that gives rise to pattern formation in our system, in contrast to others where the saturable nonlinearity dominates~\cite{PhysRevA.92.013820}. In this tight-bunching regime with $I/I_{\text{sat}}\ll1$, the effective refractive index is
\begin{equation}
n\simeq1+\frac{\chi_{\text{lin}}}{2}\left[1+\eta_1^\pm\right]=n_{\text{lin}}+n_{\text{NL}},
\label{effectiveindex}
\end{equation}
where $\chi_{\text{lin}}=-6\pi c^3(2\Delta/\Gamma)n_aC^2/[\omega_{eg}^3(1+4\Delta^2/\Gamma^2)]$ is the linear susceptibility~\cite{SchmittbergerPatternTheory}. Here, the intensity-independent terms are the linear refractive index $n_{\text{lin}}$.

The threshold for the pattern-forming instability occurs when the nonlinear optical phase shift $kn_{\text{NL}}L\gtrsim\pi/2$~\cite{PhysRevLett.17.1158}, which is achievable at low intensities using our long atomic sample and sub-Doppler temperatures~\cite{PhysRevA.90.013813}. The minimum observed threshold for pattern formation in our system is $I/I_{\text{sat}}\simeq10^{-2}$ for $\Delta=-4\Gamma$ ($\eta_1^{\pm}\simeq0.7$ and $|\zeta|\simeq2$) and an optical depth of ${\sim}50$, where $kn_{\text{NL}}L\approx1.6$. This threshold intensity is two orders of magnitude smaller than for optical pattern formation in a warm vapor~\cite{Dawes29042005}. Our typical experimental parameters are $10^{-2}\le I/I_{\text{sat}}\le 0.4$, where $\eta_1^{\pm}\simeq0.7-0.99$ and $|\zeta|\simeq2-120$. From Eq.~\ref{densitydist}, the peak density at each 1D lattice site at threshold is $\sim5n_a$, and the width of each pancake is ${\sim}\lambda/13$~\cite{PhysRevLett.69.49}. This indicates that the atoms are cooled substantially and tightly bunched in the applied 1D optical lattice, as depicted in Fig.~\ref{multimodepapersetupfig4}(a), in order to generate the instability that gives rise to pattern formation. We also note that we only observe pattern formation for $\Delta<0$ because $\Delta>0$ gives rise to reduced light-atom interaction strengths in the tight-bunching regime~\cite{PhysRevA.90.013813}.

The instability triggers a wave-mixing process that generates new, frequency-degenerate optical fields, which propagate at a small angle $\theta\sim3-10\text{ mrad}$ relative to the applied fields~\cite{Greenberg:11}, as dictated by a phase-matching condition~\cite{PhysRevLett.17.1158}. In theory, the generated fields arise anywhere along a cone centered on the $\hat{z}$-axis, as depicted in Fig.~\ref{multimodepapersetupfig4}(b). In the transverse plane, we observe multi-spot optical patterns shown in Fig.~\ref{multimodepapersetupfig4}(c) rather than a continuous ring, which indicates a spontaneous breaking of a continuous symmetry~\cite{PhysRevA.50.3471}. In addition, we observe different patterns under essentially the same experimental conditions, as shown in Fig.~\ref{multimodepapersetupfig4}(c). Such shot-to-shot fluctuations are a hallmark of non-equilibrium phenomena, where symmetry-breaking results in self-organization into different modes~\cite{PhysRevLett.91.203001,BaumannSelfOrganization}. In our instability-driven system, we observe that these fluctuations occur in as quickly as $50~\mu\text{s}$, which is on the order of the time it takes for an atom to move to a neighboring pancake and thus contribute to exciting a different pattern~\cite{SchmittbergerPatternTheory}. 

\begin{figure}
\begin{center}
 \includegraphics[scale=0.4]{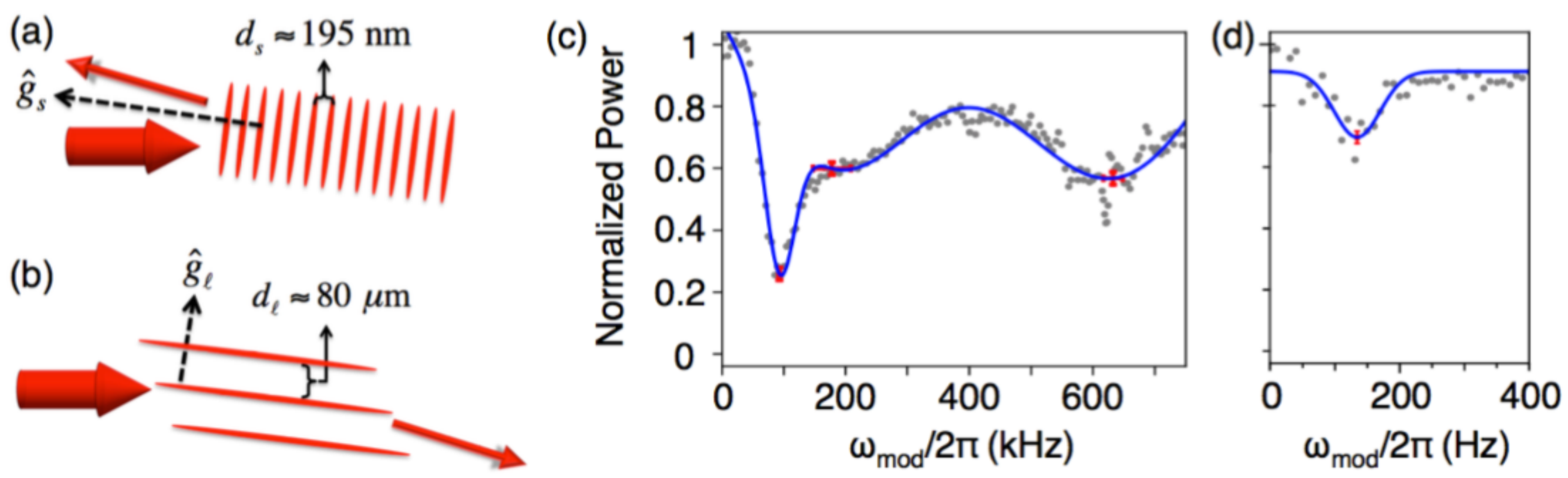}
 \caption{Example atomic patterns that form due to the interference between an applied optical field and (a) nearly counterpropagating and (b) nearly co-propagating generated optical fields. The dashed lines show the grating wavevector directions. Power of the generated light normalized to the power in the unperturbed ($\omega_{\text{mod}}/2\pi=0$) case when (c) phase-modulating one pump beam by $\pm\pi/37$ and (d) intensity-modulating an elliptical probe beam of peak intensity $I/I_{\text{sat}}=10^{-3}$, width ${\sim}1.2\text{ mm}$, and length ${\sim}4\text{ cm}$ applied to the side of the cloud (along $-\hat{r}$). Experimental data is shown in gray circles, and the best fit to a sum of Lorentzians is the blue curve with confidence intervals indicated by the error bars. The modulation depth of the short-period grating resonance is larger than that of the pump-pump resonance because more atoms are driven out of the shallower dipole potential wells for a given phase shift.}
 \label{selfgeneratedpatterns}
  \end{center}
 \end{figure}

The synergistic coupling between the optical patterns and the atoms implies that there are also corresponding real-space patterns of bunched atoms that form spontaneously with the optical patterns and are enhanced as the power in the optical patterns increases. There are two types of atomic patterns: one with a short (sub-wavelength) period, and one with a long period. The short-period gratings (spacing $d_s\simeq\pi/[2k^\prime\text{cos}(\theta/2)]\approx195\text{ nm}$) overlap strongly with the imposed pump-pump grating (spacing $d_{p}=\pi/2k^\prime$) and arise due to the interference of a generated optical field with a nearly counterpropagating pump field. The long-period gratings (spacing $d_{\ell}\simeq\pi/[2k^\prime\text{sin}(\theta/2)]\approx80\text{ }\mu\text{m}$) arise due to the interference of a generated optical field with a nearly co-propagating pump field, as depicted in Fig.~\ref{selfgeneratedpatterns}(b).

The total (applied and generated) electric field for a two-spot optical pattern is denoted by $\vec{E}(z,r,t)=[\vec{E}_0(z)+\vec{E}_w(z,r)]e^{-i\omega t}+\text{c.c.}$, where $\vec{E}_w(z,r)=i{f}_\pm(z,r)e^{ik(z\text{cos}\theta\pm r\text{sin}\theta)}\hat{y}+{b}_\pm(z,r)e^{ik(-z\text{cos}\theta\pm r\text{sin}\theta)}\hat{x}$. Here, we ignore higher-order spatial-mode patterns for simplicity. With these additional field terms, Eqs.~\ref{dipolepotentialterms} and~\ref{densitydist} are also modified so that the density distribution incorporates the self-organized atomic patterns, with corresponding grating wavevectors. The short-period grating wavevectors, $\vec{g}_s\simeq k^\prime\left\{\pm[1+\text{cos}(\theta)]\hat{z}\pm\text{sin}(\theta)\hat{r}\right\}$, are nearly along $\pm\hat{z}$, and the long-period grating wavevectors, $\vec{g}_{\ell}\simeq k^\prime\left\{\pm[1-\text{cos}(\theta)]\hat{z}\pm\text{sin}(\theta)\hat{r}\right\}$, are nearly along $\hat{r}$. The Fourier coefficients for the self-organized atomic density gratings are analogous to Eq.~\ref{FCSisyphus}, but with $\zeta\rightarrow\zeta_{s,\ell}=U_{\text{dip},(s,\ell)}/k_BT_{{s,\ell}}$, where $U_{\text{dip},(s,\ell)}$ refers to the effective dipole potential of the short-period $(s)$ and long-period $(\ell)$ gratings, and $T_{s,\ell}$ refers to the atomic temperature along $\hat{g}_{s,\ell}$. Both $U_{\text{dip},(s,\ell)}$ and $T_{{s,\ell}}$ are measured experimentally, as described below.

\section{Multimode Self-Organization}
To directly observe and characterize real-space atomic self-organization, we perform a parametric resonance experiment to measure the motional states of the atoms. The motional frequencies of atoms in the lowest-energy bound states oscillate according to
\begin{equation}
 \omega_{\text{vib},(p,s,\ell)}\simeq\sqrt{\frac{\pi^2U_{\text{dip},(p,s,\ell)}}{2md_{(p,s,\ell)}^2}},
\label{parametricresonance}
\end{equation}
where $m$ is the atomic mass, and we assume the potentials are nearly harmonic. By parametrically driving at the resonance $2\omega_{\text{vib}}$ along the direction of the grating wavevectors, atoms can be excited out of their dipole potential wells~\cite{PhysRevLett.78.2928}, which reduces the efficiency of the wave-mixing process and the power in the generated fields. Because $\hat{g}_{\ell}$ is nearly orthogonal to $\hat{g}_{p}$ and $\hat{g}_{s}$, we use different experimental methods to parametrically drive atoms in each type of grating.

To excite atoms in the short-period grating, we use an electro-optic phase modulator placed in the path of one of the pump fields and driven at frequency $\omega_{mod}$. During this experiment, we slightly misalign the pump beams to observe a stationary two-spot optical pattern and avoid detection errors due to pattern rotation~\cite{SchmittbergerPatternTheory}. We observe three distinct resonances, as seen in Fig.~\ref{selfgeneratedpatterns}(c). The high-frequency resonance at $\omega_{mod}/2\pi\simeq632\pm30\text{ kHz}$ (width $258\pm26\text{ kHz}$) corresponds to atoms trapped in the pump-pump (applied) lattice. We expect this resonance at $2\omega_{vib}/2\pi\simeq~686(+123/-109)\text{ kHz}$ according to Eq.~\ref{parametricresonance} for $U_{\text{dip},p}\equiv U_0$ with experimental parameters $\Delta=-2\pi(28\pm2)\text{ MHz}$ and $I/I_{\text{sat}}=0.28\pm0.06$. The predicted resonances, the error, and the intermediate resonance at $\omega_{mod}/2\pi=177\pm15\text{ kHz}$ are discussed further below. Using Eq.~\ref{parametricresonance}, we find that the pump-pump dipole potential well depth is $U_{\text{dip},p}\sim635\text{ }\mu\text{K}$, where the energy is normalized by $k_B/2$. Thus, the atoms at $T_z=2-3~\mu\text{K}$ are tightly confined in the imposed dipole potential wells.

The low-frequency resonance at $\omega_{mod}/2\pi=92\pm1\text{ kHz}$ (width $33\pm2\text{ kHz}$) corresponds to atoms in the short-period self-organized gratings, which we expect to occur at $\omega_{mod}/2\pi=95~(+12/-9)\text{ kHz}$ for a measured generated field intensity of $(17\pm6)\text{ }\mu\text{W/cm}^2$. The effective dipole potential well depth for atoms in the short-period gratings is therefore$U_{\text{dip},s}\sim13\text{ }\mu\text{K}$. This measurement together with the temperature measurement described below indicate that the atoms self-organize into the short-period structures, which are not imposed on the atoms by the applied fields. For the 6-spot pattern shown in Fig.~\ref{multimodepapersetupfig4}(c), we estimate there are ${\sim}7$ short-period grating lattice sites per pancake with ${\sim}20$ atoms per site, as portrayed in Fig.~\ref{multimodepapersetupfig4}(d).

To investigate the motional properties of atoms in the long-period gratings, we apply a weak elliptically-shaped optical field to the side of the atomic cloud and periodically modulate its amplitude. To detect the long-period, low-frequency resonances, we operate the experiment in the steady-state regime, where we load the MOT for 97 ms and then leave the MOT beams on at $15\%$ of their initial intensity while we run experiments. In this regime, the patterns persist for ${\sim}2\text{ sec}$, \textit{c.f.} $2.4\text{ ms}$ in the transient regime, where the MOT beams are shut off completely~\cite{SchmittbergerPatternTheory}. Here, we observe a parametric resonance at $\omega_{mod}/2\pi=134\pm2\text{ Hz}$ [Fig.~\ref{selfgeneratedpatterns}(d)] (width $50\pm7\text{ Hz}$), which we expect to occur at $\omega_{mod}/2\pi=191~(+78/-62)\text{ Hz}$. From this measurement, we find that $U_{\text{dip},l}=8\pm4\text{ }\mu\text{K}$. Because $\hat{g}_{\ell}\approx\hat{r}$, one might expect $T_{{l}}\approx T_{\text{rad}}$, which would imply $\zeta_{{l}}<1$ and negligible bunching into the long-period gratings. However, as we show below, the atoms undergo additional, spontaneous Sisyphus cooling along $\hat{g}_{\ell}$, which facilitates atomic self-organization into the long-period gratings. We therefore observe atomic self-organization not only into the sub-wavelength, short-period gratings, but also into the super-wavelength, long-period gratings.

\subsection{Analysis of the parametric resonances}
To predict the parametric resonances expected in this experiment, we calculate the effective intensity that generates the imposed and self-organized dipole potentials. The intensity of a single pump field is $I_p\sim32\pm7\text{ mW/cm}^2$, where the error is due to the measured beam size ($100\pm15\text{ }\mu\text{m}$) after the pump-beam-reshaping effect shown in Fig.~\ref{multimodepapersetupfig4}(c). This value of $I_p$ also accounts for the $10\%$ reduction in pump power from $11.2\text{ }\mu\text{W}$ to $10\text{ }\mu\text{W}$ that results from the pump-beam reshaping, discussed further below. We measure the intensity of the generated fields by measuring their output power and predicting their near-field size based on a calibration of our imaging system. We find the output intensity of a single generated field is $\sim17\pm6~\mu\text{W/cm}^2$, where the error is due to a slight asymmetry in beam size between the two spots in the optical pattern. However, because the wave-mixing process gives rise to an exponential increase in the generated field intensity across the length of the atomic cloud, we take the effective intensity inside the cloud to be the approximate intensity at the center of the cloud\textemdash$18\%$ of the output intensity, or $I_g=(3\pm1)\times10^{-3}\text{ mW/cm}^2$.

We apply these effective intensities to $U_{\text{dip},(p,s,\ell)}=\hbar\Delta I/I_{\text{sat}}^0(1+4\Delta^2/\Gamma^2)$ in Eq.~\ref{parametricresonance} with the experimental parameter $\Delta=-2\pi(28\pm2)\text{ MHz}$ and the resonant saturation intensity $I_{\text{sat}}^0=1.3\text{ mW/cm}^2$. For the pump-pump (imposed) gratings, $I=I_p$. For the self-generated gratings, $I=2\sqrt{I_p}\sqrt{I_g}$, where the factor of 2 is included because there exist two sets of self-generated gratings everywhere, \textit{e.g.}, the interference between $F(z)e^{ikz}$ and $b_+(z,r)e^{ik(-z\text{cos}\theta-r\text{sin}\theta)}$ and that between $B(z)e^{-ikz}$ and $f_-(z,r)e^{ik(z\text{cos}\theta+r\text{sin}\theta)}$ give rise to spatially overlapping dipole potentials. For this experiment, we also take $C^2\rightarrow1$ because we operate well above threshold where the fields only act on atoms that are tightly bunched in regions of pure $\hat{\sigma}^\pm$ polarizations and thus only give rise to stretched-state transitions.

Based on this analysis, we predict the parametric resonances are $2\omega_{\text{vib},s}=686~(+123/-109)\text{ kHz}$, $2\omega_{\text{vib},s}=95~(+12/-9)\text{ kHz}$, and $2\omega_{\text{vib},\ell}=191~(+78/-62)\text{ Hz}$, where the larger error for the long-period resonance is due to the error in $\theta=4\pm1\text{ mrad}$. These predicted values agree with the measured values of $2\omega_{\text{vib},p}=632\pm30\text{ kHz}$, $2\omega_{\text{vib},s}=92\pm1\text{ kHz}$, and $2\omega_{\text{vib},\ell}=134\pm2\text{ Hz}$ to within the experimental error.

For these self-organized atoms, the location of the parametric resonance is a function of both $U_{\text{dip,p}}$ and $n$; \textit{e.g.}, it decreases with lower pump beam intensities. Overall, we observe that the short-period-grating parametric resonance occurs between $90\text{ and }133\text{ kHz}$ and the long-period-grating parametric resonance occurs between $11\text{ and }134\text{ Hz}$ for $I/I_{\text{sat}}=0.05\text{ to }0.3$.

We also attribute the parametric resonance at $\omega_{mod}/2\pi=177\pm15\text{ kHz}$ (width $213\pm34\text{ kHz}$) to the dipole potential that arises due to the small ring around the pump beams, appearing in Fig.~1(c). This ring is pump-beam reshaping effect that arises because the pump size is comparable to the width of the cloud of atoms. We model this ring as an LG$_{10}$ mode and find it contains $\sim10\%$ of the power contained in the central pump spot. The peak intensity of the ring is therefore $3.6\%$ of the pump intensity. We predict that the dipole potentials generated by this ring and a nearly counterpropagating pump should have a parametric resonance at $172~(+38/-27)\text{ kHz}$, which agrees with our measured value of $177\pm15\text{ kHz}$. In this prediction, we neglect the dipole potentials due to a ring and a nearly copropagating pump field because the lattice spacing for these are $\sim250\text{ }\mu\text{m}$, and thus there only exist 0 or 1 such gratings in the cloud of atoms.

We attribute the large width of the pump-pump resonance to the anharmonicity of the potential well, which is characterized by a modified ground state energy of the form $E_0=\hbar\omega_{\text{vib},p}\left(1-A_1/2\right)/2$, where $A_1$ is known as the first-order anharmonicity parameter~\cite{Fletcher2002}. If we consider the deviation of the mean measured predicted resonance ($2\omega_{\text{vib},p}=632\text{ kHz}$) from the mean of the theoretically predicted value assuming a harmonic potential ($2\omega_{\text{vib},s}=686\text{ kHz}$), we find $A_1=0.15$. This anharmonicity gives rise to a reduction in the overall frequency and a broadening of the parametric resonance, in agreement with previous work~\cite{PhysRevLett.78.2928}.

With this analysis, we find that our data agrees well with theory, and we thus demonstrate real-space self-organization of atoms in a multimode system.

\section{Bragg scattering}
We also perform a Bragg scattering experiment to both determine the temperature of the self-organized atoms~\cite{Mitsunaga:98} and provide an additional characterization of their real-space structure. We apply the counterpropagating pump fields and allow the patterns to form and persist for $200~\mu$s. We then shut off the pump fields and inject a weak probe beam along either the $\pm\hat{z}$-direction. We only collect light from a portion of one of the emission cones, as depicted in Fig.~\ref{multimodepapersetupfig4}(b). In this case, a probe beam traveling along $-\hat{z}$ ($+\hat{z}$) is Bragg-matched to scatter into multiple directions, but it will only reach the detector if it back-scatters (forward-scatters) off the short-period (long-period) gratings depicted in Fig.~\ref{selfgeneratedpatterns}(a),(b). This technique provides further evidence of real-space self-organization and allows us to distinguish between the two scales of long-range ordering.
\begin{figure}
\begin{center}
 \includegraphics[scale=0.34]{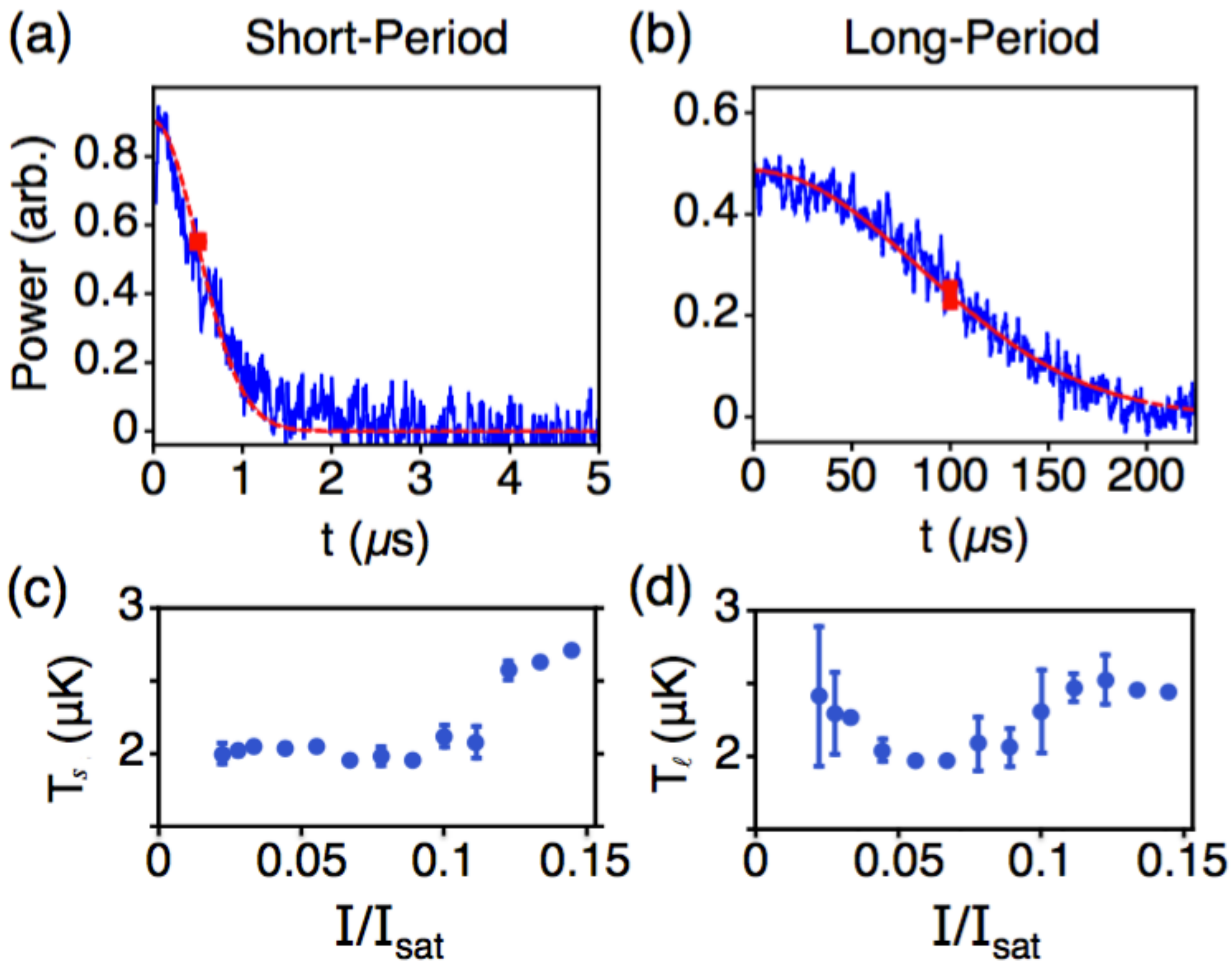}
\caption{Temporal evolution of the scattered probe power. Pattern formation occurs during $-200\text{ }\mu\text{s}\le t\le-65\text{ ns}$ with pump beam intensities $I/I_{\text{sat}}=0.1$. A probe beam is turned on at $t=0$ in the (a) $-\hat{z}$-direction and (b) $+\hat{z}$-direction. The red (dashed) curve is a best fit to a Gaussian, with (a) $\tau_s=0.9\pm0.02\text{ }\mu\text{s}$ and (b) $\tau_{\ell}=120\pm2\text{ }\mu\text{s}$. The rectangular error bars at (a) $t=0.5\text{ }\mu\text{s}$ and (b) $t=100\text{ }\mu\text{s}$ indicate the typical statistical confidence interval of the fit. The temperature of atoms in the (c) short-period and (d) long-period self-organized gratings are extracted from data similar to (a) and (b) for various $I/I_{\text{sat}}$. The error bars define the statistical standard deviation.}
 \label{Braggscattering}
  \end{center}
 \end{figure}

Bragg scattering allows us to extract the temperature of atoms in the short- and long-period gratings because ballistic expansion of atoms out of their gratings results in the decay of the probe signal~\cite{Mitsunaga:98}, as shown in Fig.~\ref{Braggscattering}. We note that the atoms move ballistically, rather than diffusively, once released from the optical lattice because the mean free path in our system ($\approx23$ mm) is much longer than any of the lattice constants. Ballistic expansion is modeled using a Gaussian decay function $f_{s,\ell}(t)=a\text{exp}(-t^2/\tau_{s,\ell}^2)$~\cite{Mitsunaga:98}, where $\tau_{s,\ell}$ corresponds to the time it takes for $n_{\text{NL}}$ to reduce to $1/e$ of its initial value. Because $d_{\ell}\gg d_s$, the peak density in the long-period gratings decays more slowly and consequently $\tau_{\ell}\gg \tau_s$.

\subsection{Analysis of the atomic temperatures}
To extract an atomic temperature from the Bragg scattering data shown in Figs.~\ref{Braggscattering}(a) and (b), we use a heuristic model of the atomic density distribution to relate $\tau_{s,\ell}$ to $T_{s,\ell}$. The density distribution for the short-period gratings before ballistic expansion is approximately
\begin{equation}
\eta_{s}(z,r)\approx\eta_{s}(z)\approx\frac{n_a}{2I_0(-\zeta_{s})}\left[e^{-\zeta_{s}\text{cos}(2k^\prime_w z)}+e^{-\zeta_{s}\text{cos}(2k^\prime_w z-\pi)}\right],
\end{equation}
where we use the experimentally measured value of $U_{\text{dip,s}}$ in $\zeta_{s}$. We note that $k^\prime_w$ is slightly larger than $k^\prime$ because the pattern-forming optical fields experience a different index of refraction~\cite{PhysRevLett.17.1158,SchmittbergerPatternTheory}. However, for the distance scales of relevance in this problem, it is a good approximation to take $k^\prime_w\approx k^\prime$.

\begin{figure}
\begin{center}
 \includegraphics[scale=0.36]{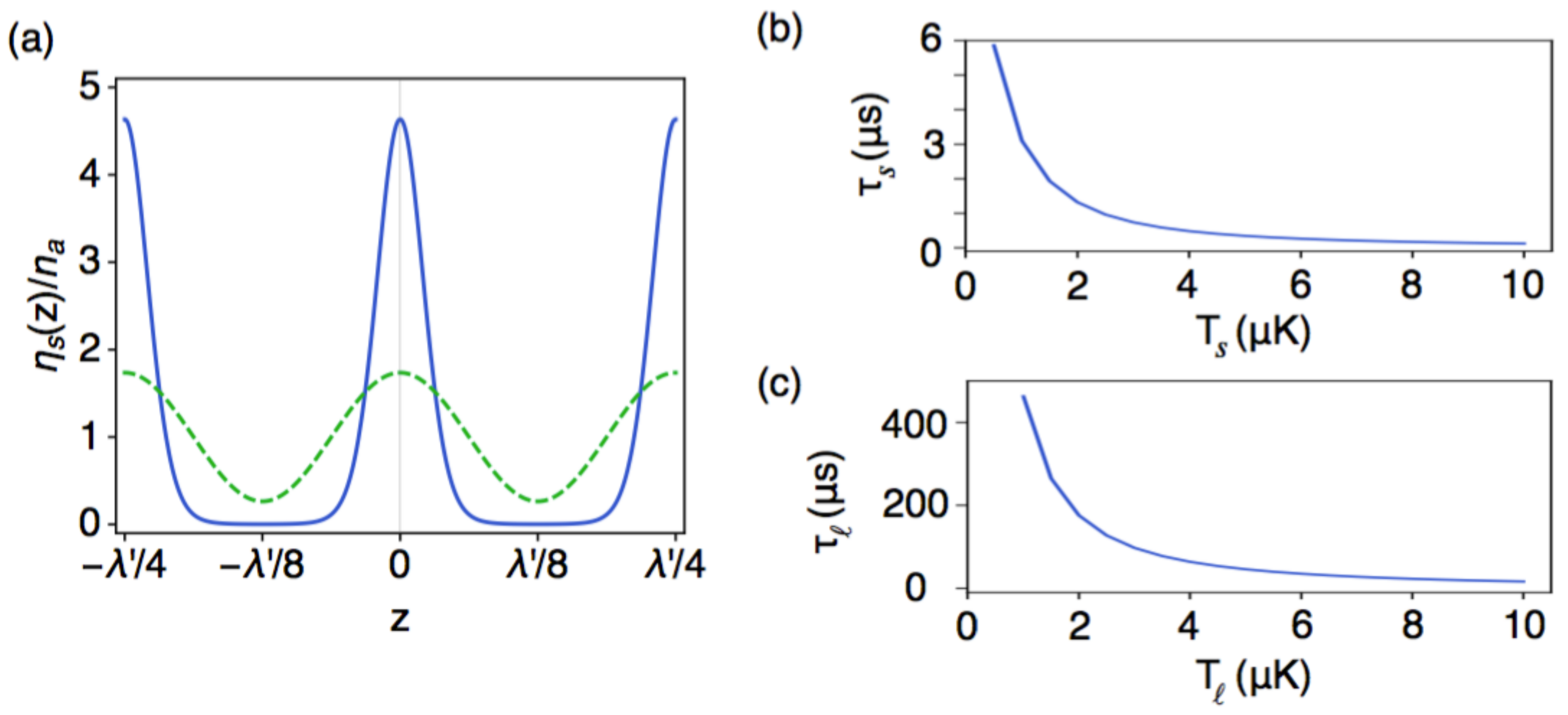}
 \caption{\textbf{Modeling ballistic expansion.} (a) The blue (solid) curve shows the ``before'' density distribution $\eta_s(z)$ for example parameters $I/I_{\text{sat}}=0.01$, $\Delta/\Gamma=-5$, and $T/T_D=3/146$. The green (dashed) curve shows the ``after'' density distribution $\eta^\prime_s(z)$ where the peak density is reduced by $1/e$. The predicted decay constant $\tau$ as a function of $T$ for example parameters $\tilde{\Delta}=-5$ and $I_{\text{eff}}=0.7$ $\text{mW}/\text{cm}^2$ in the case of the (b) short-period gratings and (c) long-period gratings.}
 \label{heuristictempmodel}
  \end{center}
 \end{figure}

The density distribution after ballistic expansion for the short-period gratings can be approximated as $\eta^\prime_{s}(z)\approx n_a\left[1+f(t)\text{cos}(4k^\prime z)\right]$ in the case where $f<1$. The constraint $f<1$ maintains normalization, and it is valid for our experimental regime. We calculate the magnitude of $|f(t=\tau_s)|$ for a given $\eta_{s}(z)$ such that the peak density of $\eta^\prime_{s}(z)$ is $1/e$ that of $\eta_s(z)$. Example density distributions at $t=0$ and $t=\tau_s$ are shown together in Fig.~\ref{heuristictempmodel}(a) in blue (solid) and green (dashed), respectively. 
 
We fit the ``before'' and ``after'' density distributions between $\pm\lambda^\prime/8$ to a Gaussian envelope in order to calculate the ``before'' and ``after'' grating widths. We apply this as the characteristic distance $d$ in a simple kinematics model of $v_0=d/\tau$, where $v_0=\sqrt{3k_BT/m}$. Leaving $T_s$ as an unknown parameter in $\zeta_{s}$, we find a numerical solution for the temperature as a function of $\tau_s$. An example of this predicted relationship is shown in Fig.~\ref{heuristictempmodel}(b) for $\tilde{\Delta}=-5$ and $I/I_{\text{sat}}=0.01$. Using the same procedure, we relate $\tau_l$ and $T_l$ for the long-period gratings, whose results are shown in Fig.~\ref{heuristictempmodel}(c).

From these relationships, we extract an atomic temperature from our experimentally measured decay constants of the Bragg-scattered probe signal. We perform multiple Bragg scattering experiments to extract $\tau_s$ and $\tau_{\ell}$ for various pump intensities, which generates the data shown in Figs.~\ref{Braggscattering}(c) and (d). Figure~\ref{Braggscattering}(c)/(d) shows the atomic temperature is $T_{{s}}\approx1.8-2.8\text{ }\mu$K along $\hat{g}_s$ and $T_{\ell}\approx1.9-2.7\text{ }\mu$K along $\hat{g}_{\ell}$ for all pump intensities. We interpret the slight upward trend as a result of the fact that Sisyphus cooling operates optimally over a limited range of intensities~\cite{SisyphusCastinmain}.

The low temperatures along $\hat{g}_s$ are expected because atoms in these gratings substantially overlap with the imposed gratings and therefore have the advantage of undergoing initial Sisyphus cooling due to the applied fields. However, we also observe that atoms in the long-period gratings cool to comparable temperatures despite having a momentum nearly orthogonal to $\hat{z}$.

Such transverse cooling can arise due to a weak transverse dipole force in a 1D optical lattice~\cite{PhysRevA.80.052703}, Raman cooling~\cite{PhysRevLett.69.1741}, or due to 3D Sisyphus cooling that arises above threshold for pattern formation due to the interaction of the generated fields and the pump fields. However, the weak transverse dipole force imposed by the 1D lattice does not give rise to efficient cooling, \textit{i.e.}, it would take approximately $600\text{ }\mu\text{s}$ for the atoms to move a distance $d_{\ell}$ under the influence of this weak force alone. In contrast, we have observed the signature of these long-period gratings as soon as $20\text{ }\mu\text{s}$ after turning on the pump beams, and thus, this force cannot be responsible for the cooling timescales that we observe. In addition, we have never observed Raman transitions to other ground states, and thus we can only attribute the observed cooling to frequency-degenerate schemes such as Sisyphus cooling.

We thus conclude that the only mechanism that can cool and trap atoms in the long-period gratings so quickly and effectively is Sisyphus cooling, where the expected damping time is $[\hbar k^2|\Delta|/(2m\Gamma)]^{-1}\approx10\text{ }\mu\text{s}$~\cite{Metcalf}. The observed 3D Sisyphus cooling process occurs spontaneously as a result of the optical/atomic pattern-forming instability. Our observation of 3D cooling by only applying fields along one dimension allows us to achieve longer coherence times with a simplified geometry in comparison to lattice experiments where laser beams are applied in all three dimensions~\cite{PhysRevLett.78.630}. This is also supported by our observation that the patterns persist in the transient regime for up to $2.4~\text{ms}$~\cite{SchmittbergerPatternTheory} \textit{cf.} ${\sim}300~\mu\text{s}$ in similar wave-mixing experiments where 3D cooling is absent~\cite{PhysRevA.86.013823}, thus giving rise to more rapid atom loss.

\section{Conclusion}
In conclusion, we directly measure the real-space self-organization of atoms in a multimode geometry using parametric driving and Bragg scattering microscopy, and we observe spontaneous three-dimensional Sisyphus cooling. Our system exhibits continuous symmetry-breaking and long-range atomic self-structuring on multiple spatial scales. Our work represents an important step towards studying non-equilibrium phenomena in multimode geometries and provides a simplified system in which one can observe low-light-level, multidimensional nonlinear optical effects.

\section*{Acknowledgements}
We gratefully acknowledge the support of the National Science Foundation through Grant $\#$PHY-1206040 as well as helpful discussions with Dr.~Thorsten Ackemann and Dr.~Joel Greenberg.

\bibliography{refsmultimodepaper}

\end{document}